\documentclass[11pt]{article}

\usepackage{lipsum} 
\usepackage{graphicx}

\usepackage[sc]{mathpazo} 
\usepackage[T1]{fontenc} 
\usepackage{microtype} 

\usepackage[hmarginratio=1:1,top=32mm,columnsep=20pt]{geometry} 
\usepackage{multicol} 
\usepackage[hang, small,labelfont=bf,up,textfont=it,up]{caption} 
\usepackage{booktabs} 
\usepackage{float} 
\usepackage{hyperref} 

\usepackage{lettrine} 
\usepackage{paralist} 

\usepackage[english]{babel} 

\usepackage{abstract} 

\usepackage{titlesec} 
\titleformat{\section}[block]{\large\scshape\centering}{\thesection.}{1em}{} 
\titleformat{\subsection}[block]{\large}{\thesubsection.}{1em}{} 

\usepackage{fancyhdr} 
\title{\vspace{-15mm}\fontsize{24pt}{10pt}\selectfont\textbf{Hybrid Update/Invalidate Schemes for Cache Coherence Protocols}} 

\author{
\large
\textsc{Roman Dovgopol\textsuperscript{1} and Matthew Rosonke\textsuperscript{2}}\\ 
\\
\normalsize \textsuperscript{1} Kaspersky Lab, Moscow, Russia\\
\normalsize \textsuperscript{2} University of Minnesota --- Twin Cities, Minneapolis, USA\\ 
\vspace{-5mm}
}
\date{}

\begin{document}

\maketitle 

\thispagestyle{fancy} 



\section*{Abstract}

\textit{In general when considering cache coherence, write back schemes are the default. These schemes invalidate all other copies of a data block during a write. In this paper we propose several hybrid schemes that will switch between updating and invalidating on processor writes at runtime, depending on program conditions. We created our own cache simulator on which we could implement our schemes, and generated data sets from both commercial benchmarks and through artificial methods to run on the simulator. We analyze the results of running the benchmarks with various schemes, and suggest further research that can be done in this area.}

\begin{multicols}{2}

\section{Introduction}
\noindent When the first microprocessor was released, its memory operations were relatively short when compared to their corresponding arithmetic operations. Since then, microprocessors have been trending strongly in the other directions, with today's load and store operations being several orders of magnitude slower than arithmetic operations. This so called 'memory wall' has only been exacerbated by the coming of microprocessors. The added complexity of trying to synchronize memory operations and, more importantly, cache contents between cores can tremendously slow down performance if not executed intelligently. In this paper, we will discuss variations of the standard MOESI cache coherence scheme that allow a cache to either update or invalidate during a write request, depending on the situation.\\

\subsection{Background}

\noindent The most common and widely used state-based coherence scheme in multi-core machines is the MOESI scheme. It consists of the following five states:\\

\noindent \textit{(M)odified} --- The cache block is the sole owner of 'dirty' data.\\
\noindent \textit{(O)wned} ---- The cache block owns the 'dirty' data, but there are other sharers.  A cache with a block in the O state processes requests for that block from other cores.\\
\noindent \textit{(E)xclusive} ---- The cache is the sole owner of clean data.\\
\noindent (\textit{S)hared} ---- The cache is one of several possessors of a block, but it is not the owner and its data is clean.\\
\noindent \textit{(I)nvalid} ---- The cache block does not hold valid data\\

\noindent In general, most machines will use an Invalidate protocol with NMOESI.  That is, when two caches contain blocks with the same tag, a write to one cache causes an invalidation signal to be sent to the other cache.  A cache will send out this invalidate signal unless it knows it is the sole owner of the data, such as in the M or E state.  In the case of the O, S or I state, the cache will generate and invalidate signal that will tell all other caches to set their copies of the data to I.\\

\noindent Invalidate schemes can be thought of a reactive approach to cache coherence.  A cache will only receive modified data from another cache if it asks for it.  For a more proactive approach, one would look to an update scheme.  \\

\noindent An update signal is sent with data in the same scenarios where an invalidate scheme would send an invalidate signal, but rather than set their blocks to I, these cores would replace their old data with the block's new value and set it to the S state.  
Both schemes have their advantages and disadvantages.  It's good to be proactive and use and update scheme if you know that a block written to by one core will soon be read by another core, but updates can also generate a lot of unnecessary bus traffic. \\ 

\noindent Meanwhile, invalidate schemes will avoid this bus traffic up front, but may still generate it later if they need to read a block that has been invalidated.  Like most things, \textit{it is possible that a good answer lies somewhere in between}.  Below, we propose hybrid schemes that switch between invalidating and updating depending on the cores' recent behavior.\\

\subsection{Previous Research}

\noindent A fair amount of research was done on the advantages and disadvantages of updating or invalidating in the mid-80s.  Since then, most research has gone towards other aspects of coherence, but many of these papers present a reasonable starting place.\\

\noindent A method called the RB protocol was proposed by Rudolf and Segall \cite{PS84} for write-through caches.  The scheme updates all other cores on the write-through by default, but if two writes occurred back to back, data in all other cores would be invalidated.  This likely saved traffic for write-through machines, but as most machines today have write-back caches, updating on every write would create an excessive amount of extra bus traffic.\\

\noindent Karlin, Manasse, Rudolf and Sleator \cite{KM86} would later propose a scheme called 'Competitive Snooping' which would rely on amortized analysis to allow updates to occur so long as there was enough allotted cost for them to occur.  This cost was related to the amount of time it would have taken if invalidation had occurred instead, but that invalidation eventually resulted in cache misses. While interesting, this scheme would likely also struggle on write-back machines.  As we will show later, it is much better to invalidate by default and update when necessary.\\

\noindent While both above methods relied mainly on the patterns of their own cores, Archibald \cite{AJ88} proposed a scheme that would take into account the actions of other cores. Once again, it updated by default, but if any core had three writes to a single location without any other core accessing that location, invalidation would occur instead. We also see a potential profit of hybrid schemes in various fields such as large-scale systems with shared memory \cite{SM11}\cite{HB11}, memory-optimized protocols \cite{LM06}, and others.

\noindent Our proposed schemes all begin by invalidating first, then allowing updates when certain criteria have been met.  They also heavily take into account the actions of other cores on the network.\\

\section{Proposed Schemes}

\noindent For our research, we decided to implement and compare several different schemes for performance:\\

\subsection{Invalidate-Only Scheme}

\noindent This is the basic scheme that is used by many multicore systems. When a cache writes to a block in the O, S or I state, it sends an invalidate signal to the network. All other cores that receive this signal invalidate their copies of the block.\\

\subsection{Update-Only Scheme}
\noindent The opposite of the Invalidate-Only Scheme, caches writing to a block in the O, S or I state send an update signal with data to the network.  All other cores that receive this signal update their copies with the correct value and set themselves to S.\\

\subsection{Threshold Scheme}

\noindent This is the first of our proposed hybrid schemes that we implemented ourselves.  In this scheme, each cache block carries with it an associated counter that is used to determine whether updates or invalidates should occur upon a write.  It is defined by the following three scenarios:\\
\begin{enumerate}
	\item Upon entry to the cache from main memory, counter is initialized to zero.
	\item Whenever a read request is seen by a cache and it contains a valid block with matching address, that block's counter is increased by one.
	\item After a block is successfully written to, its value decreases by one.
\end{enumerate}

\noindent When we write to a block, we check the counter value against the threshold. If the counter is above or equal to the threshold, we send an update signal to the network.  Otherwise, we send an invalidate signal. The logic behind this scheme is two-fold.  When we sense multiple reads to a block, we increase the counter and aim to update rather than invalidate. When we sense more writes, we have a lower counter and invalidate other blocks instead.\\

\subsection{Adapted-MOESI}

\noindent This scheme is the same as the Invalidate-Only scheme except that when writing to a block that is in the O state, we send an update signal to the network rather than an invalidate signal.  Invalidation still occurs when writing to a block in the S or I state. As we will discuss later, the Threshold Scheme works best with a threshold of one. When a block's counter is set to one, its state is almost always zero, so this scheme attempts to approximate the effects of the threshold scheme without the extra hardware.\\

\subsection{Number of Sharers Scheme}

\noindent Our final scheme is an alternate version of the threshold scheme. Rather than keep track of read and write requests to a memory location, whether or not to do an update is determined by the number of sharers any given data block has. If the number is above or equal to a certain number of sharers, an update will occur in place of an invalidate. This is particularly relevant due to its ease of implementation in directory schemes, whose popularity is on the rise in highly parallel machines.\\

\section{Simulation}

\subsection{Creating the Simulator}

\noindent In order to simulate each of these different schemes, our team developed a simple cache simulating program in C++.  The program takes as input a list of loads and stores, with each string in the list containing a load/store identifier, a core number, and an address. When run with one of these inputs, the program simulates the operation of anywhere from 1 to 16 separate caches under the standard MOESI protocol.  During the run, it keeps track of the number of reads, writes, read request, write requests (invalidates) and update requests at each core.  Since our program simulates the scheme functionality independent of timing, we are looking at the total number of read requests, write requests and update requests as our metric for performance. The total number of requests is proportional to the amount of traffic that would exist on the network and therefore is an acceptable means of judging performance.
We chose to develop our own simulator mainly for speed of simulation and ease of programming. Doing so gave us the freedom to keep track of whatever metrics we liked, while also being able to easily add in various different versions of the coherence scheme.  Other simulators like multi2sim, which is discussed in the next section, proved to be incredibly difficult to make changes to and were significantly slower due to all of the additional work that goes into the full timing simulation. Ultimately, it was decided that timing simulation was less important than the functional simulation, since timing varies so greatly from machine to machine.\\

\subsection{Simulation Statistics}

\noindent Our simulator can simulate anywhere from 2 to 16 caches at once. The simulator only uses one level of caches.  Beyond the first level, all caches are connected to main memory.  Each cache contains 64 sets with 4 blocks in each set.  Each dataset that we generated to run on the simulator contains roughly five million loads/stores, so the metric used in this paper will be the total number of read requests, invalidates and updates on all cores per five million instructions.\\

\section{Generating Datasets}

\noindent In order to run our simulator, we needed to generate files containing list of loads and stores to the various cores.  We chose to look at a diverse array of datasets in order to gain the best possible understanding of our various schemes.  Also, we made sure that generated datasets are reasonably representative of their benchmark.\\

\noindent Each of these benchmarks was run on 2, 4, 8 and 16 cores.  Each scenario was simulated \textit{using Invalidate-Only, Update-Only, Threshold, Adapted-MOESI and Number of Sharers} schemes.\\

\subsection{Commercial Workloads}

\noindent We certainly wanted to include datasets corresponding to commercial benchmarks. To do this, we took advantage of the multi2sim timing simulator \cite{MU14}.  While it was very difficult to implement the new hybrid schemes in the multi2sim timing simulator, we found that it was easy to adapt the simulator to generate datasets. While running a timing simulation, we had the simulator output to a file the information for five million consecutive loads and stores. We usually waited several tens of millions of instructions for the parallel programs to get warmed up before starting the output. This way, we were able to generate a more representative sample of the benchmark's performance. We generated datasets from the following four benchmarks in this way.\\

\noindent \textit{Bodytrack} --- Computer vision algorithm\\
\noindent \textit{Dedup} --- Compression of a data stream through local and global means\\
\noindent \textit{Streamcluster} --- Solves online clustering problem\\
\noindent \textit{Swaptions} --- uses Monte Carlo techniques to price a portfolio of swaptions\\

\subsection{Artificial Workloads}

\noindent Finally, we created a handful of pseudo-random datasets meant to represent common multicore scenarios, such as many cores sharing a lock, many cores updating an array based on an element's neighbors, and a server model. These datasets were generated with simple C++ programs.\\

\noindent Our \textit{Locks} dataset established 3 shared locks between any number of cores. Each core had a 10\% chance of accessing the lock.  When doing so, the core would write to the lock to free it if it possessed it. If it did not possess the lock, it would read from the lock and then write to take the lock if no one else possessed it.  Only blocks containing the locks were shared between cores.  All other data accesses were restricted to their own private range of addresses.\\

\noindent Our \textit{Arrays} dataset represents an array that is constantly updated by comparing elements. In this scenario, an array element is read by one core, as are its neighbors above, below to the right and to the left of it. Each core traversed through a row in this array, and during each cycle, a core would be randomly chosen to process the next element in its row. Note that in a real program, this would result in non-deterministic behavior.\\

\noindent Our \textit{Pseudo-Server} dataset represents a very basic server-client model with public and private data where one core is allowed to write to shared data and each other core may only read from it.  The server core can write to any block in the whole address range.  The address range itself is split into two sections.  The first section is public and can be read by any client core.  The second section represents private space and is divided between all of the client cores which are only allowed to read from their own space.\\

\section{Results and Analysis}

\noindent Below we present results and analysis for each scheme using the various benchmarks. Note that all graphs only display the total sum of all bus transactions for each scenario.  Detailed breakdown of how those transactions are split between read requests, invalidates and updates is provided in the appendix.\\

\subsection{Invalidate/Update Only Scheme}

\noindent First, we will simply look at the base \textit{Invalidate-Only} and \textit{Update-Only} schemes. To limit the amount of data presented in this section, only graphs for 8-core scenarios are presented, although results from scenarios with other numbers of cores will be discussed. Additionally, as mentioned above, the numbers presented are bus transactions per five million memory instructions. Results for the commercial and artificial workloads are shown below (\textbf{Figure 1}).\\

\noindent The primary point gained from this data is that, for many applications, there is a large gap between the number of transactions that occur with an update-only scheme and an invalidate-only scheme. In many workloads, the amount of data that is heavily shared between cores is much less than the amount of data that is primarily used by one core but is occasionally accessed by others. In an update-only scheme, we are updating any core that has ever accessed the shared data, when we ideally only want to update those cores that have accessed it recently.
The one exception to this pattern is the \textit{bodytrack} benchmark.  The difference between the two schemes is relatively small, indicating denser sharing between the caches. As we will see later, this makes this benchmark a good candidate to improve performance under a hybrid scheme (\textbf{Figure 2}).\\

\noindent Our artificially generated benchmarks present much less variation between the two extremes.  The pseudo-server benchmark, due to its unique structure, actually performs better under the update-only scheme.\\

\noindent Another interesting note to take away is that the \textit{arrays} benchmark maintains a consistent number of transactions regardless of scheme, even though the distribution of updates/invalidates is different. Due to the 'enforced' order of the memory transactions (they happen in order on each core, although the core that may proceed in each iteration is chosen randomly), the benchmark never really benefits from any updates.\\

\subsection{Threshold and Adapted-MOESI Schemes}

\noindent In this section, we will analyze the results from running the benchmarks with the \textit{Threshold} scheme at several different thresholds, as well as under the \textit{Adapted-MOESI scheme} (\textbf{Figure 3}). \\

\begin{figure*}[!htb]
\minipage{0.32\textwidth}
  \includegraphics[width=\linewidth]{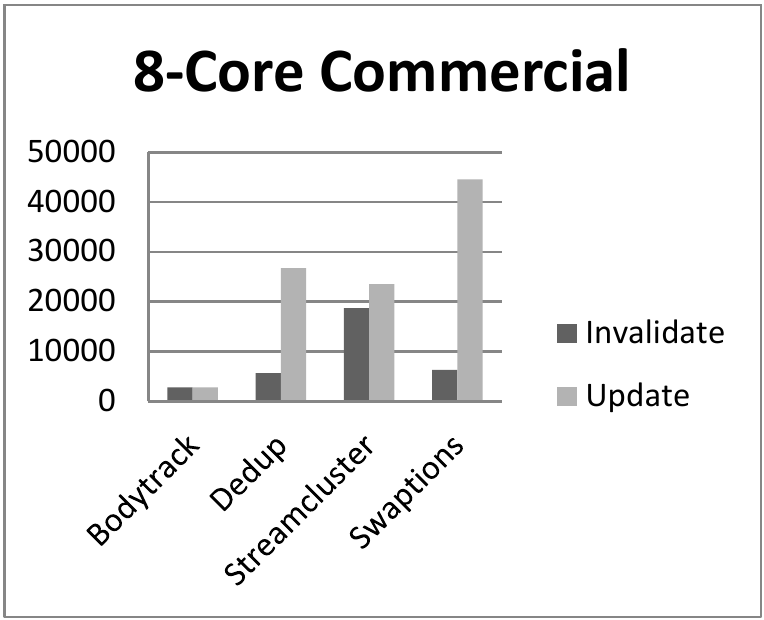}
  \caption{A really Awesome Image}\label{fig:awesome_image1}
\endminipage\hfill
\minipage{0.32\textwidth}
  \includegraphics[width=\linewidth]{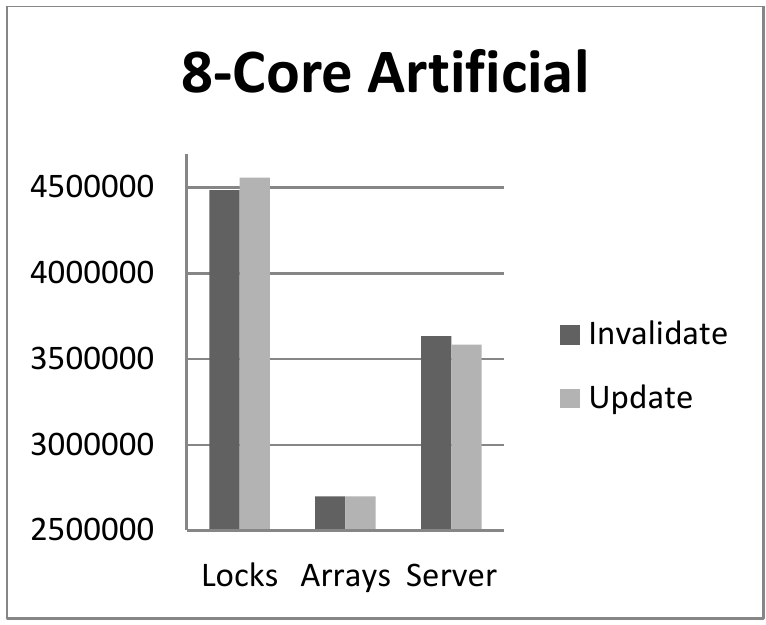}
  \caption{A really Awesome Image}\label{fig:awesome_image2}
\endminipage\hfill
\minipage{0.32\textwidth}%
  \includegraphics[width=\linewidth]{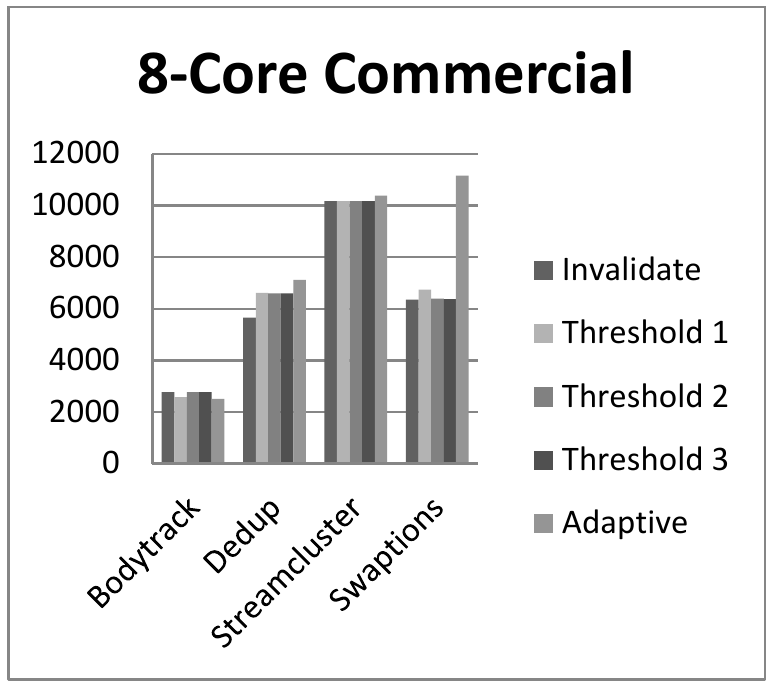}
  \caption{A really Awesome Image}\label{fig:awesome_image3}
\endminipage
\end{figure*}

\noindent For the most part, there is a much smaller gap between the number of transactions that occur with the \textit{Invalidate-Only} scheme and the hybrid scheme. Still, for those benchmarks that originally had a large gap, the \textit{Invalidate-Only} scheme outperforms any hybrid scheme. For \textit{bodytrack}, however, the hybrid schemes of \textit{Threshold 1} and \textit{Adapted-MOESI} actually outperform the other schemes.  Since the benchmark was relatively dense, and because the update and invalidate schemes both performed relatively well, having a smart way to choose whether to update or invalidate ends up improving performance.\\

\noindent When it came to the value to set the \textit{Threshold} to, only a value of one really showed any difference from an Invalidate-Only scheme. The \textit{Threshold} of 3 was in most cases identical to running \textit{with Invalidate-Only}.  \\

\noindent Due to this result, we believed that it may be worthwhile to implement a scheme that updates when the state of the block being written to was (O)wned. This logic stemmed from the observation that when the threshold of one was met, the block was most commonly in the O state. In practice, however, this performed not better that a \textit{Threshold} of one, but at times would perform significantly worse.  While blocks with a counter value that met the threshold of one were often in the O state, not all blocks in the O state would necessarily have a threshold value of one (\textbf{Figure 4}).\\

\noindent It is somewhat difficult to tell because of the scale of the graph, but the \textit{locks} benchmark performed slightly worse with the \textit{Threshold } scheme than it did with the \textit{Invalidate-Only} scheme, while the server benchmark did slightly better.  The arrays benchmark still did not see any change.\\

\noindent The server benchmark is interesting because it was the only one to do better under the \textit{Update-Only} scheme.  In this case, the \textit{Threshold } and \textit{Adapted-MOESI} schemes did better than always invalidating, but worse than always updating.  While these hybrid schemes will not necessarily be the best possible scheme for each benchmark, they may provide a decent compromise between schemes that perform best always invalidating and those that perform best always updating.\\

\begin{figure*}[!htb]
\minipage{0.32\textwidth}
  \includegraphics[width=\linewidth]{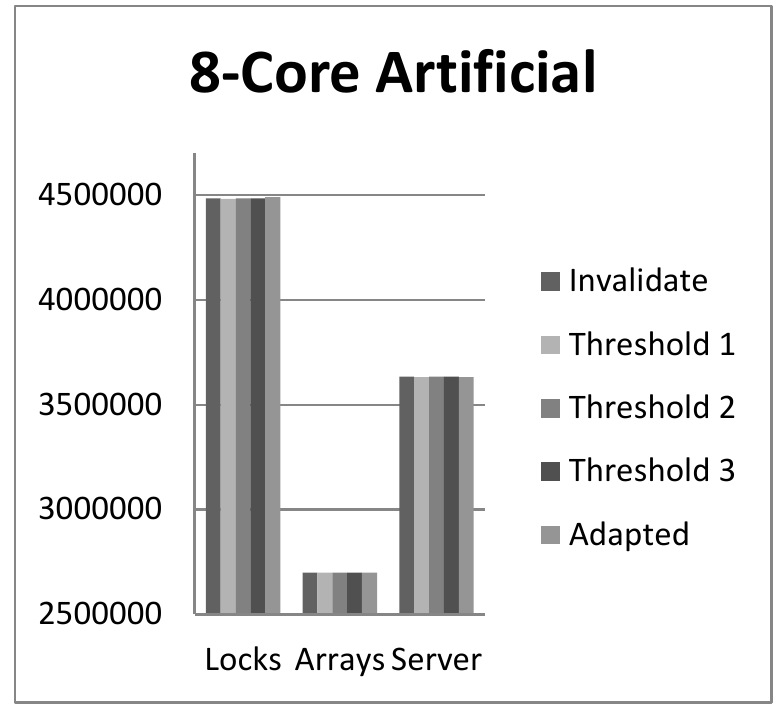}
  \caption{A really Awesome Image}\label{fig:awesome_image1}
\endminipage\hfill
\minipage{0.32\textwidth}
  \includegraphics[width=\linewidth]{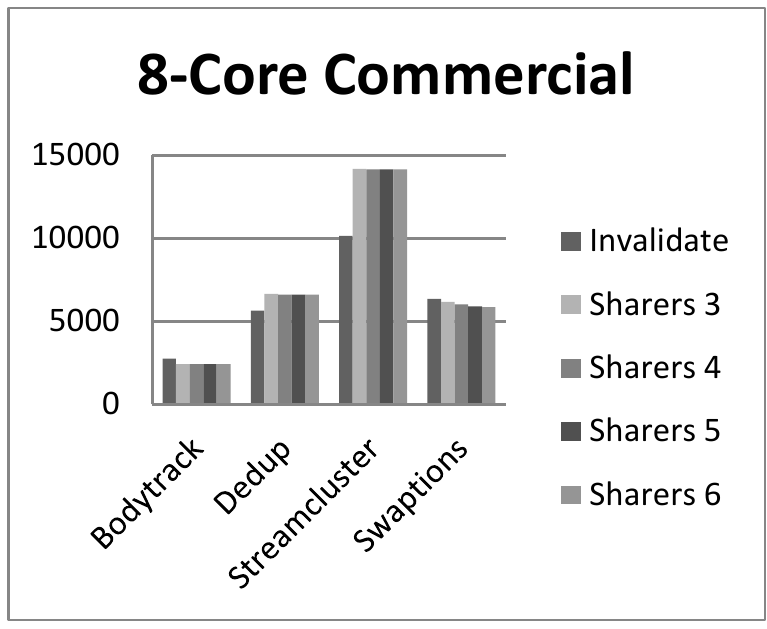}
  \caption{A really Awesome Image}\label{fig:awesome_image2}
\endminipage\hfill
\minipage{0.32\textwidth}%
  \includegraphics[width=\linewidth]{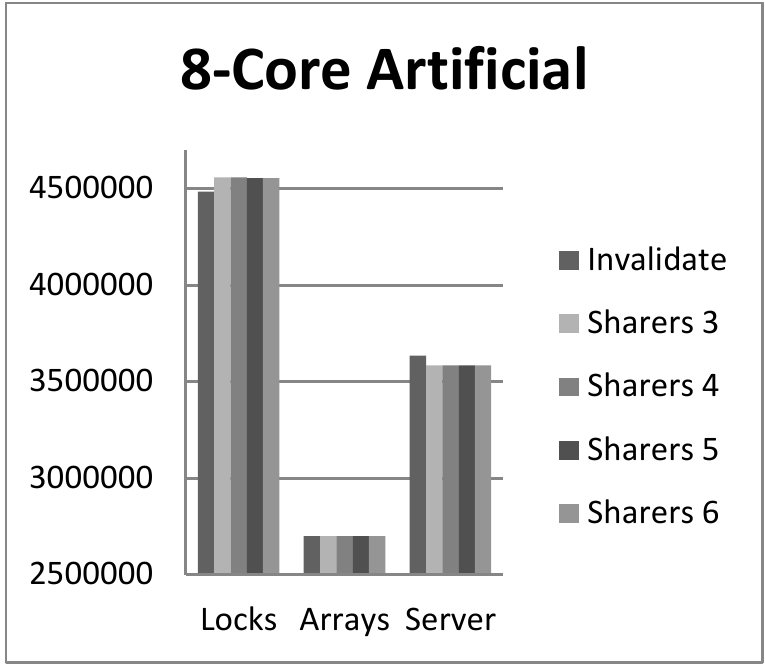}
  \caption{A really Awesome Image}\label{fig:awesome_image3}
\endminipage
\end{figure*}

\subsection{Number of Sharers Scheme}

\noindent Finally, we will address the results gained from running each benchmark under the \textit{Number of Sharers} scheme (\textbf{Figure 5}).\\

\noindent The \textit{Number of Sharers} scheme actually performs relatively well in most cases. Like the \textit{Threshold} scheme, it performs better on the \textit{bodytrack} benchmark than either always updating or always invalidating.  Interestingly, the \textit{swaptions} benchmark also sees improvement. Unlike the \textit{Threshold} scheme, this scheme has the benefit of always knowing exactly how many other caches share data with a cache that is being written to, and this seems to be reflected as an increase in performance on some benchmarks.\\

\noindent On other benchmarks, specifically \textit{streamcluster}, this scheme seems to perform worse.  Because of how the updating works, the only way for a core not to become a sharer again is to be evicted from the cache, since it will never be invalidated once updates start happening. If a core doesn't access a block regularly but also doesn't evict it often enough, the scheme may update when it doesn't need to. This effect is reflected in the poor performance of the \textit{streamcluster} benchmark (\textbf{Figure 6}). \\

\noindent Finally, the results for the \textit{Number of Sharers} scheme on the artificial benchmarks look very similar to the \textit{Threshold scheme}, except the results are more exaggerated. It does worse on the \textit{locks} benchmark but better on the \textit{server} benchmark. Because of the factors discussed above, this scheme seems to be more of a win-more/lose-more scheme than the \textit{Threshold} scheme. If a benchmark benefitted from the \textit{Threshold} scheme relative to the \textit{Invalidate-Only} scheme, it benefits more with the Number of Sharers scheme. If it did worse with \textit{Threshold}, it does even worse with Number of Sharers.\\

\noindent The minimum number of sharers required for updates to occur seemed to be best set around half of the number of cores.  If it was too little, such as two sharers in the case of eight cores, then too many updates occurred. When the required number of sharers got above half, the performance usually stagnated at a constant value, since anything that is shared between half of the cores is generally shared between almost all of them.\\

\section{Final Points}

\noindent In this final section of the paper, we will discuss what conclusions can be drawn from the above analyzed data, what additional considerations need to be taken into account when judging the results, and suggest further research that can be done in this area.\\

\subsection{Conclusions}
\noindent There certainly exist examples of benchmarks that perform better with either an \textit{Invalidate-Only} scheme or an \textit{Update-Only} scheme. In some instances, such as the bodytrack benchmark, there exist hybrids that perform better than either \textit{Invalidate-Only} or \textit{Update-Only}. In other instances, there are hybrid schemes that will perform better than one of \textit{Invalidate-Only} or \textit{Update-Only} but worse than the other.  \\

\noindent When considering different threshold values for the \textit{Threshold } scheme, a value of 1 provided the most dramatic result.  High threshold values functioned almost identically to \textit{Invalidate-Only} schemes. Employing the \textit{Threshold} scheme with a value of one resulted in the lowest number of transactions on some benchmarks, while providing a reasonable compromise on others.\\

\noindent The Adapted-MOESI scheme did not perform as well as expected, as it led to more bus transactions than the Threshold scheme in every scenario.\\

\noindent Finally, the \textit{Number of Sharers} scheme performed reasonably well, especially when the required number of sharers needed to perform an update was around half the number of cores. However, it varied more from the average than the Threshold scheme did. Because of this, the Threshold scheme seems to be the correct choice for a scheme that will provide the optimal compromise between benchmarks that perform best with more updates and those that perform best with more invalidates.\\

\subsection{Additional Considerations}

\noindent Our simulator did not take timing into account, as we were only concerned with counting the total number of transactions.  Since the timing would vary from machine to machine, metrics such as IPC would be less informative than the total number of transactions.  In a real machine, the timing of updates and invalidates plays an important role. Updating results in longer stores but potentially much faster loads, while invalidation can do the reverse.\\

\noindent We also did not consider hardware cost when evaluating the various schemes. Updating on its own requires more hardware since more complex transactions must be sent over the bus.  The \textit{Threshold} scheme requires substantial extra hardware, since each cache block must contain its own counter. The \textit{Adapted-MOESI} scheme requires virtually no extra hardware. The Number of Sharers requires some sort of centralized index of the number of sharers on all data blocks in all caches. This can be easily accomplished by the directory in any cache coherence protocol that uses one.\\

\subsection{Further Research}

\noindent While the \textit{Adapted-MOESI} scheme was meant to emulate a \textit{Threshold} scheme with a threshold value of one using less hardware, it ultimately failed in that endeavor. Still there is certainly a way to get the same effect with significantly less hardware.\\

\noindent While we chose not to concern ourselves with the timing effects of the various schemes, they would certainly be interesting to address.\\

\noindent Finally, since our simulator used a snoopy protocol combined with \textit{MOESI}, it would be interesting to see how each of these schemes interacts with a directory-based protocol.  It would be especially interesting for the \textit{Number of Sharers} scheme, as that scheme would be so easy to implement in a directory-protocol.\\



\end{multicols}


\end{document}